\documentclass[12pt]{iopart}
\usepackage{iopams}
\usepackage{graphicx}
\usepackage{color}

\begin{document}

\newcommand{\be}{\begin{equation}}
\newcommand{\ee}{\end{equation}}
\newcommand{\bea}{\begin{eqnarray}}
\newcommand{\eea}{\end{eqnarray}}
\newcommand{\nn}{\nonumber}
\newcommand{\de}{\partial}

% Greek alphabet
\def\a{\alpha}
\def\b{\beta}
\def\d{\delta}        \def\D{\Delta}
\def\e{\epsilon}
\def\eps{\varepsilon}
\def\f{\phi}          \def\F{\Phi}
\def\vf{\varphi}
\def\g{\gamma}        \def\G{\Gamma}
\def\h{\eta}
\def\i{\iota}
\def\j{\psi}          \def\J{\Psi}
\def\k{\kappa}
\def\lam{\lambda}   \def\L{\Lambda}
\def\m{\mu}
\def\n{\nu}
\def\o{\omega}   \def\O{\Omega}
\def\p{\pi}      \def\P{\Pi}
\def\q{\theta}   \def\Q{\Theta}
\def\r{\rho}
\def\s{\sigma}   \def\S{\Sigma}
\def\t{\tau}
\def\u{\upsilon}  \def\U{\Upsilon}
\def\x{\xi}      \def\X{\Xi}
\def\z{\zeta}

\title[Resonant VE electron--CO cross sections]{Electron--impact resonant vibration excitation cross sections and rate coefficients for carbon monoxide}

\author{
V. Laporta$^{1,a}$\footnote[0]{$^a$ v.laporta@ucl.ac.uk}, C. M. Cassidy$^{1}$, J. Tennyson$^{1,b}$\footnote[0]{$^b$ j.tennyson@ucl.ac.uk} and R. Celiberto$^{2,3,c}$\footnote[0]{$^c$ r.celiberto@poliba.it}
}

\address{$^1$ Department of Physics and Astronomy, University College London, London WC1E 6BT, UK}
\address{$^2$ Department of Water Engineering and Chemistry, Polytechnic of Bari, 70125 Bari, Italy}
\address{$^3$ Institute of Inorganic Methodologies and Plasmas, CNR, 70125 Bari, Italy}

\begin{abstract}
Resonant vibrational and rotation-vibration excitation cross sections for electron--CO scattering are calculated in the 0--10~eV energy range for all 81 vibrational states of CO, assuming that the excitation occur via the $^2\Pi$ shape resonance. Static exchange plus polarization calculations performed using the R-matrix method are used to estimate resonance positions and widths as functions of internuclear separation. The effects of nuclear motion are considered using a local complex potential model. Good agreement is obtained with available experimental data on excitation from the vibrational ground state. Excitation rates and cross sections are provided as a functions of the initial CO vibrational state for all ground state vibrational levels.
\end{abstract}

\maketitle

\section{Introduction}

When a spacecraft enters an atmosphere at speeds exceeding the local speed of sound, a shock wave is formed behind it and its kinetic energy is transformed into heat. The energy delivered to the gas in this process promotes excitation of the molecular internal degrees of freedom (rotational, vibrational and electronic) and chemical reactions (including dissociation and ionization). The hot reacting (and radiating) gas is often in thermal and chemical nonequilibrium conditions and it is this complex system that interacts with the vehicle surface.

In this paper we consider electron-impact resonant vibrational excitation and rotational-vibrational excitation of carbon monoxide over the entire range of vibrational levels supported by its ground electronic state, CO($X^1\Sigma^+$), and over an extended electron temperature range. The energy range of concern here is 0--10~eV. This data provides important input information into spacecraft entry models for atmospheres such as those on Mars and Venus as well as to study other processes in the atmospheres of these planets~\cite{campbell2009atm_mars_venus} and in comets~\cite{cb09}. It is also useful for understanding processes involved in the CO laser~\cite{haddad,Wang2001475} and to study CO plasma in presence of electrical discharge~\cite{Gorse1984165,Gorse1984177}. Infrared emission from CO in the upper atmospheres of Mars, Venus and several other planets is a subject of current theoretical and experimental interest. The first measurements of cross sections for low-energy electron impact excitation of the vibrational levels of the ground state of CO have been made by Schulz~\cite{PhysRev.135.A988}.  Recent new measurements showing the contribution of electron impact relative to emissions by other mechanisms, have been reported~\cite{campbell2009atm_mars_venus}.

Direct vibrational excitation of diatomic molecules by electron impact is, in general, an inefficient process because of the small electron-to-molecule mass ratio. However, when the incident electron can attach to form a temporary negative ion, vibrational excitation cross section can be hugely enhanced~\cite{0034-4885-31-2-302}. These processes are called resonant collisions. In this paper the following reaction is treated:
\begin{equation}
e^- + \textrm{CO}(X^1\Sigma^+, v_i) \to \textrm{CO}^-(^2\Pi) \to e'^-+ \textrm{CO}(X~^1\Sigma^+, v_f)\,,
\label{eq.res_coll}
\end{equation}
where $v_i$ and $v_f$ are the initial and final vibrational levels of the molecule. A number of reviews on this subject are available~\cite{RevModPhys.45.423, Trajmar1983219, Michael1989219, Brunger2002215}.

Resonance enhanced vibrational excitation of CO from its vibrational ground state ($v_i=0$) has been well-studied experimentally, notably in recent work by Allan~\cite{PhysRevA.81.042706}, Popari\ifmmode \acute{c}\else \'{c}\fi{} {\it et al.}~\cite{PhysRevA.73.062713} and by Gibson {\it et al.}~\cite{0953-4075-29-14-026}; earlier experimental work is reviewed by Brunger and Buckman~\cite{Brunger2002215}. Theoretically the best calculations are due to Morgan~\cite{0953-4075-24-21-015, 0953-4075-26-15-026} who used an R-matrix method to characterize the resonance as a function of CO internuclear distance. As will be shown below, her studies show rather good agreement with both data available to her at the time and subsequent studies. The various studies show that electron collisions involving the well-known, low-lying $^2\Pi$ shape resonance lead not only to a large vibrational excitation cross section but also show that $\Delta v$ ($=v_f - v_i$) can be large: excitation cross sections with $\Delta v =11$ were measured by Allan~\cite{PhysRevA.81.042706}.

So far studies of electron impact vibrational excitation of CO have concentrated almost exclusively on excitations from the vibrational ground state. However for modeling CO in hot environments it is necessary to know the corresponding cross sections starting from excited vibrational states,  $v_i>0$. In this work we develop a model which reproduces the known data for $v_i=0$ and then use it to obtain vibrational excitation cross sections and rates for the whole range of possible initial vibrational states.

The paper is organized as follows: In section \ref{sec:th_model} the theoretical model is presented for both electron dynamics and nuclear motion; in section \ref{sec:results} the calculated results are presented and in section \ref{sec:summary} they are discussed.

\section{Theoretical model \label{sec:th_model}}

\subsection{Electron collisions}

The R-matrix method~\cite{jt474, B2011} is used to obtain a complex potential energy curve for the resonance. Our methodology and calculations are heavily based on the work of Morgan~\cite{0953-4075-24-21-015} who, as is shown below, obtained excellent results for the excitation process. For the present study it was necessary to extend the range of Morgan's calculations to allow for excitation to high-lying bound states. To do this we follow the prescription for the electron collision calculations given in her paper but, since the paper does not provide entire details of the calculation, it was necessary to first perform a number of test studies. The nuclear dynamics is treated in the local-complex-potential model as explained in the next subsection.

Calculations were performed using the diatomic molecule implementation of the UK Molecular R-matrix codes~\cite{jt225} which uses Slater Type Orbitals (STOs) to represent the target and numerical functions to represent the continuum. We used the STO basis given by Kirby-Docken and Liu~\cite{ISI:A1977DG11400007} to generate a total of 16$\sigma$ and 12$\pi$ molecular orbitals, and numerical Bessel functions for partial waves up to $l=6$ for the continuum inside the R-matrix sphere which had a radius of 10~a$_0$. To avoid linear dependence problems, two of the continuum $\pi$ orbitals were removed by Lagrange orthogonalization to the target orbitals~\cite{jt61}. Resonance positions and widths were obtained by fitting the $^2\Pi$ eigenphase sum at geometry  to a Breit-Wigner form using program RESON \cite{jt31}. The resonance becomes very broad at short internuclear separations and, under these circumstances, the background eigenphase can vary significantly across the resonance resulting in fits that are less stable. The effect of this can be seen in the behavior of the fitted width as a function of CO bondlength.  See Fig.~\ref{gammaCO} below.

Morgan tested a number of models but found that a Static Exchange plus Polarisation (SEP) model performed best. This model uses a Hartree-Fock (HF) target wavefunction and includes polarisation effects augmenting the scattering wavefunction with so called two particle -- one hole (2p-1h) configurations. These configurations involve simultaneously exciting a single target electron into an unoccupied target (``virtual'') orbital and placing the scattering electron in a virtual orbital. Results for resonances parameters are well-known to be sensitive to the precise choice of parameters in an SEP calculation~\cite{0022-3700-17-12-022}. Our final model froze the C and O 1s and 2s electrons and considered excitation of remain 6 ``valence'' electrons into virtual orbitals, all of which were retained. At the CO equilibrium bondlength this model gives a resonance position and width of (1.67, 0.82)~eV which can be compared to Morgan's values of (1.68, 0.95)~eV. It can be seen that the positions are in excellent agreement but our width is somewhat narrower. As discussed below, it was decided to increase our calculated widths by 10~\%\ to improve agreement with experiment.

\subsection{Nuclear motion}

Our treatment of vibrational excitation by low-energy electron impact follows the general formulation given in the refs.~\cite{0034-4885-31-2-302, Domcke199197}. At low energies the cross section is dominated by negative ion resonance contributions, that is the incident electron is temporarily captured by the molecule and a negative intermediate state occurs, then the resonant state decays into a new state. In this paper, process (\ref{eq.res_coll}) is described within the framework of the \textit{local-complex-potential} (LCP) model~\cite{PhysRevA.20.194, wadehra, PhysRevA.77.012714}, that is an approximation to the more general \textit{non--local} theory, which is appropiate when the resonance width is much larger than the spacing between the target vibrational levels (for CO $\sim0.1$~eV). In general the resonance width depends on the energy as well as on the internuclear distance, $\Gamma(E,R)$; the LCP approximation replaces this energy-dependence with the value $E(R)$, the fixed-nuclei resonance energy~\cite{0022-3700-16-5-019}: \be \Gamma(E,R)\approx\Gamma(E(R),R)=\Gamma(R)\,.\ee In the following all dynamical quantities are understood energy-independent.

In the LCP approach the electron-molecule cross section, in the rest frame of the molecule for vibrational transition $v_i\to v_f$ and for an incident electron with energy $\e_i$, is given by~\cite{0034-4885-31-2-302}:
\be
\s_{if}(\e_i) = \frac{16\,\pi^4\,m_e }{\hslash^2}g\ \frac{k_f}{k_i}\ \left|\langle \chi_f|{ \cal V}|\xi\rangle\right|^2\,, \label{eq:xsec}
\ee
where $m_e$ is the electron mass and $g=2$ is a statistical factor. $k^2_{i(f)}=2\,m_e\,\e_{i(f)}/\hslash^2$ is the incoming (outgoing) electron momenta, $\xi(R)$, depending on the internuclear distance $R$, is the solution of the nuclear wave equation of the negative ion, with total energy $E=\e_i + \epsilon_{v_i}$, \be \left(T_J + V^- + \D-\frac i 2\G - E \right)\x(R) = -{\cal V}\ \chi_i(R)\,,\label{eq:wave_nucl}\ee and $\chi_{i(f)}(R)$ is the initial (final) vibrational eigenfunctions of the neutral molecule with eigenvalue $\epsilon_{v_{i(f)}}$, given by the equation: \be \left(T_J +   V_0\right)\chi_n(R) = \epsilon_{v_n}\ \chi_n(R)\,. \label{eq:eigen_neutral}\ee  The kinetic energy operator, $T_J(R)$, including the centrifugal potential, is expressed as \be T_J(R) = -\frac{\hslash^2}{2\,\mu }\frac{d^2}{dR^2} + \frac{J(J+1)\hslash^2}{2\mu R^2}\,,\ee where $\mu$ is the nuclei reduced mass and $J$ the angular momentum quantum number of the target molecule. The adiabatic potentials of the target and resonant electronic states, $V_0(R)$ and $V^-(R)$, have been represented by a Morse function: \be V(R) = D_e\left[1-e^{-\a(R-R_e)}\right]^2+W\,. \label{eq:morse_pot} \ee

In the LCP model, the discrete--state--continuum coupling potential ${\cal V}$ is expressed as: \be {\cal V}^2(R) = \frac1{2\pi}\frac{\G(R)}{k(R)}\,, \label{eq:VkR} \ee where $k(R)$ is defined by \be k^2(R) = \frac{2\,m}{\hslash^2}\left|V^--V_0\right|\,. \label{eq:k_electron} \ee The level shift operator $\Delta(R)$ and the resonance width $\Gamma(R)$ are discussed in the next section.

Once the collision cross section $\s_{if}$ and the electron energy distribution are known, the vibrational excitation rate coefficient K$_{if}$ can be  evaluate. Assuming that the electron energy distribution is Maxwellian, the rate coefficient can be expressed, as a function of the electron temperature $T$, as: \be \textrm{K}_{if}(T) = \frac{2}{\sqrt\pi}\,(\kappa T)^{-1.5}\,\int^\infty_{\e_{th}}\,\e\,\s_{if}(\e)\,e^{-\e/\kappa T}\,d\e\,, \label{rate_coeff} \ee where $\kappa$ is the Boltzmann constant and $\e_{th}=v_f-v_i$ is the threshold energy for the process $i\to f$.

\section{Results \label{sec:results}}

\begin{figure}
%\begin{center}
\begin{indented}
\item[]
\includegraphics[scale=.7,angle=0]{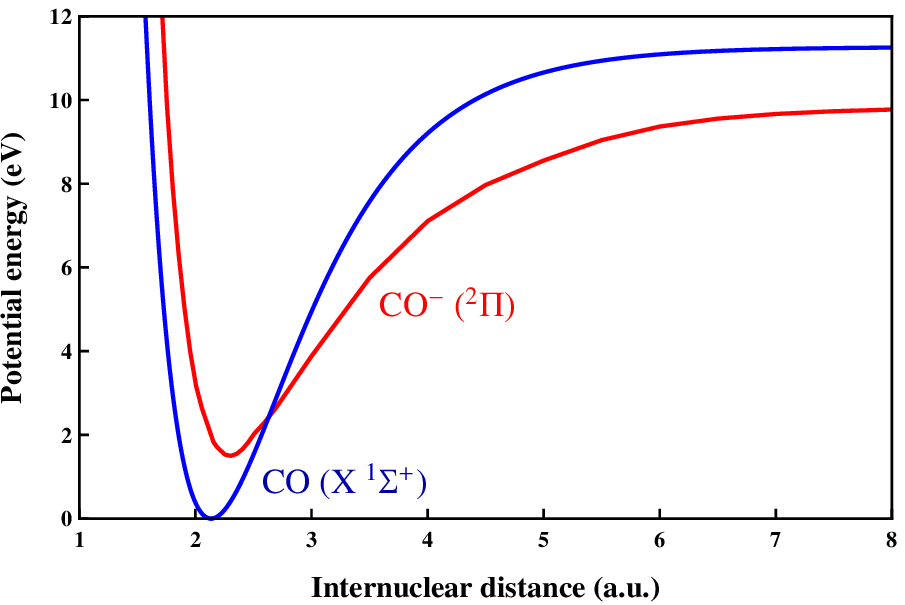}\hspace{.5cm}
\includegraphics[scale=.7,angle=0]{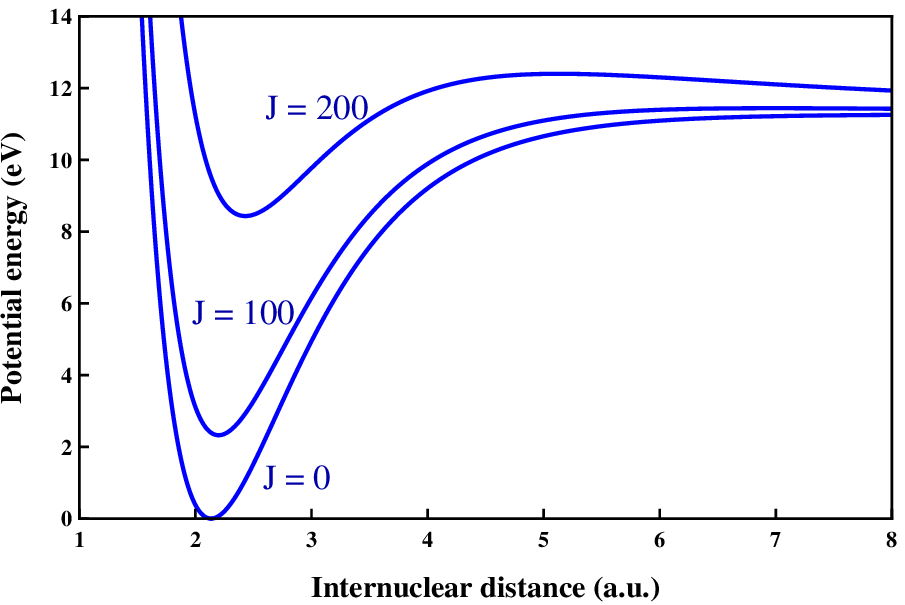}
%\end{center}
\end{indented}
\caption{Left: CO and CO$^-$ potential energy curves.  Right: CO ground state potential energy curves including the rotational contribution coming from different values of
the rotational angular momentum, $J$, as indicated in the figure. \label{morseCO}}
\end{figure}

The ground state of the potential energy curve for carbon monoxide CO(X~$^1\Sigma^+$) was calculated using Molpro~\cite{molpro2009} and an aug-cc-pwCV5Z GTO basis in a Davidson-corrected Multi-Reference Configuration Interaction (MRCI) calculation. The calculated points have been fitted with a Morse function and the parameters are displayed in Table~\ref{tab:COmorseparam}, compared with those of different authors. The CO$^-(^2\Pi)$ potential curve was taken from the R-Matrix described above up to the crossing point with the CO potential at $R=2.6$~a$_0$; at longer bondlengths a Molpro calculation was used to extend the curve to its asymptotic limit. This curve matched the resonance curve at the crossing point. The resonance potential was made to reflect the true shape of the CO potential energy curve by using resonances energies relative to our calculated target curve rather than the HF curve used in the calculations. Our curves give an asymptotic electron affinity of 1.43~eV, close to the observed oxygen atom electron affinity of 1.46~eV~\cite{PhysRevA.81.022503}. It is important to get these details correct as the vibrational excitation cross sections are very sensitive to the relative positions of the CO and CO$^-$ curves. Fig.~\ref{morseCO} shows in the left panel the obtained CO and CO$^-$ potential energy curves as a function of the internuclear distance. The CO($X~^1\Sigma^+$) ground state has been found to support 81 vibrational states. The righthand panel of Fig.~1 shows  curves representing the ground state potential energies plus the angular contribution coming from different rotational states ($J=0, J=100$, and $J=200$).

\begin{table}
\begin{center}
\begin{tabular}{r|ccc|cc}
\hline\hline
             & \multicolumn{3}{c|}{CO($X ^1\Sigma^+$)} & \multicolumn{2}{c}{CO$^-(^2\Pi)$} \\
             & ~This work~ & ~Theory~\cite{Liu20112296}~ & ~Exp.~\cite{herzberg}~ & ~This work~ & ~Theory~\cite{0953-4075-24-21-015}~ \\\hline
$D_e$ (eV)   & 11.26     & 11.19                     & 11.22               & 9.76      & -    \\
$R_e$ (a.u.) & 2.13      & 2.13                      & 2.13                & 2.30      & 2.25 \\
$W$ (eV)     & 0         & -                         & -                   & 1.49      & -    \\
\hline\hline
\end{tabular}
\caption{Dissociation energy ($D_e$), equilibrium distance ($R_e$) and 	minimum position ($W$) for the CO and CO$^-$ Morse-like parameters of Eq. (\ref{eq:morse_pot}), compared with those of different authors. \label{tab:COmorseparam}}
\end{center}
\end{table}

\begin{figure}
%\begin{center}
\begin{indented}
\item[]
\includegraphics[scale=.7,angle=0]{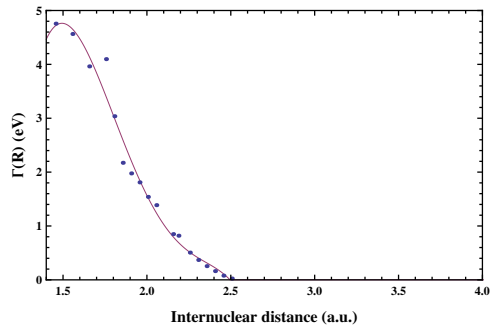}
%\end{center}
\end{indented}
\caption{Resonance width, $\G(R)$, as a function of bondlength. The figure shows the the R-matrix points and the fitting curve given by Eq. (\ref{fitting func}.)\label{gammaCO}}
\end{figure}

\begin{figure}
%\begin{center}
\begin{indented}
\item[]
\includegraphics[scale=.7,angle=0]{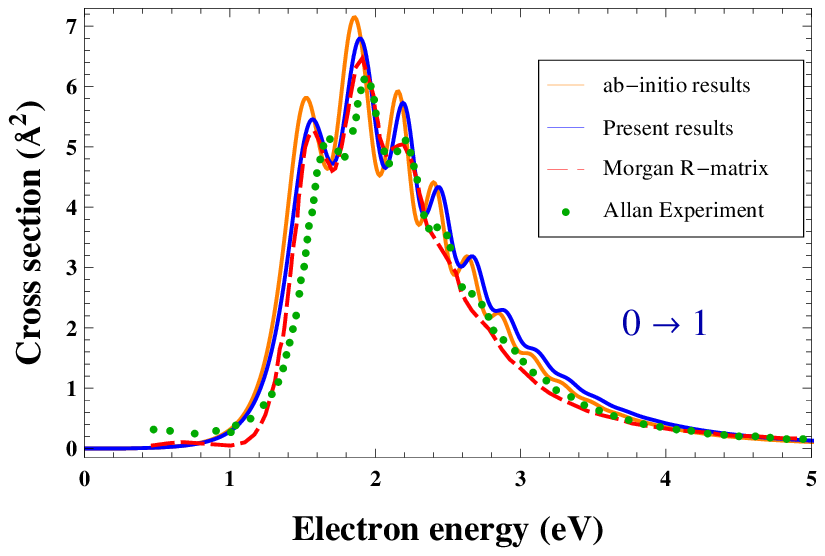}
\includegraphics[scale=.7,angle=0]{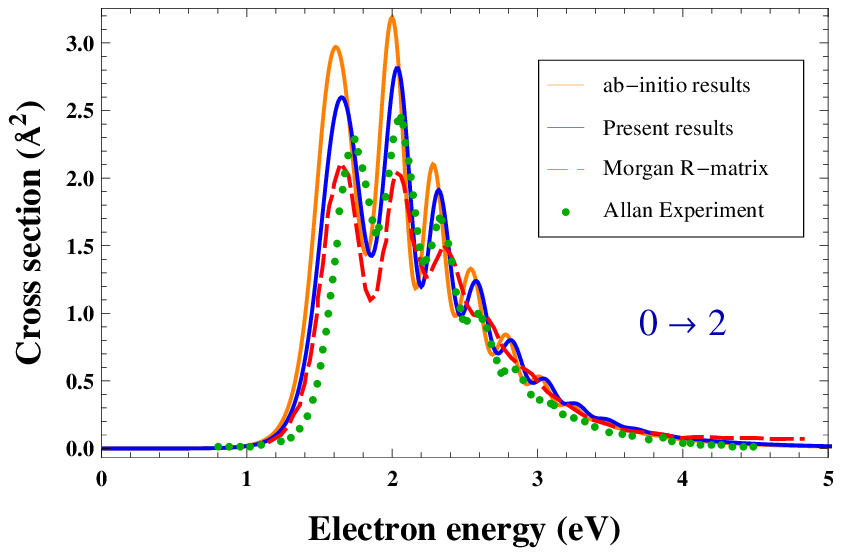}
\includegraphics[scale=.7,angle=0]{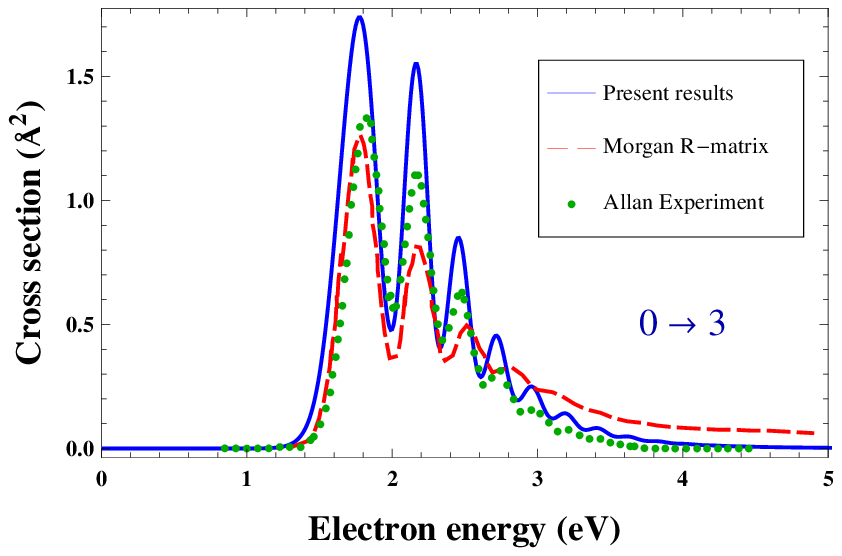}
\includegraphics[scale=.7,angle=0]{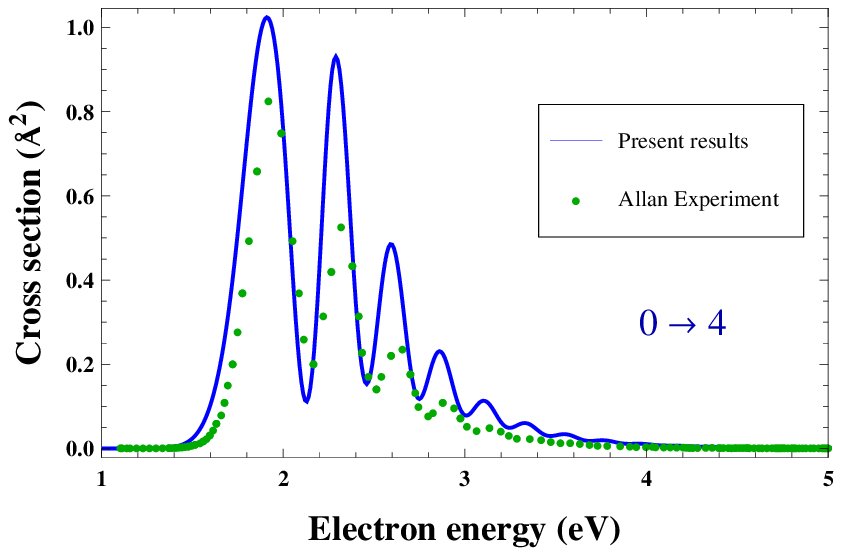}
\includegraphics[scale=.7,angle=0]{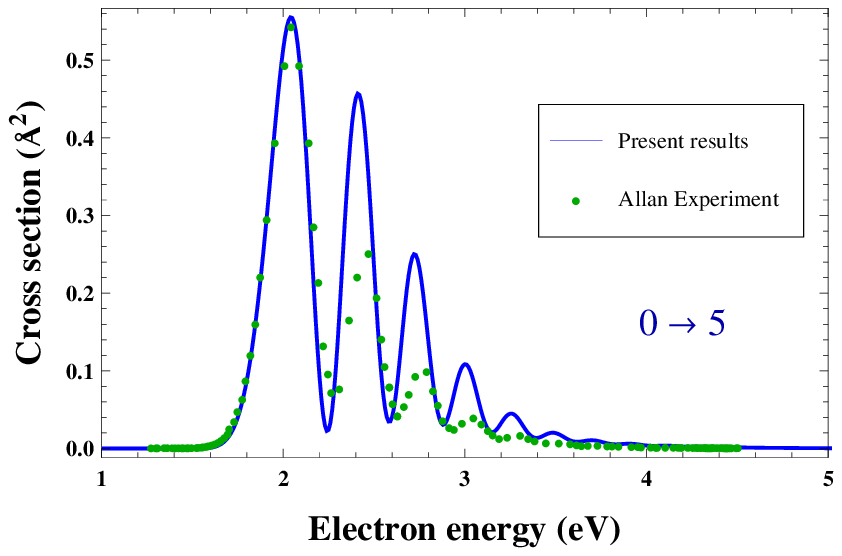}
\includegraphics[scale=.7,angle=0]{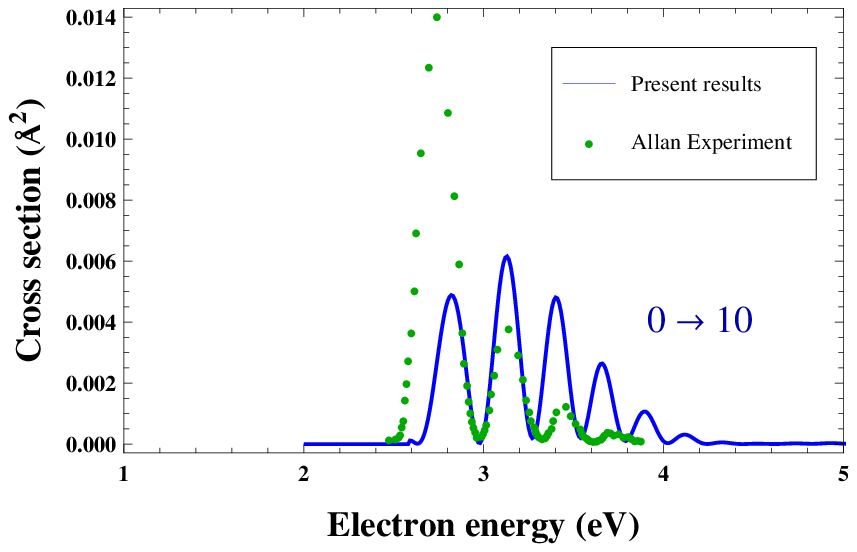}
%\end{center}
\end{indented}
\caption{e-CO resonant cross sections compared with the experimental results of Allan~\cite{PhysRevA.81.042706} and the theoretical R-matrix calculation of Morgan~\cite{0953-4075-24-21-015}. \label{fig:COxsec_allan}}
\end{figure}

The resonance width, $\G(R)$, has been fitted with a polynomial function,
\be
\G(R)= (c_1 + c_2 R + c_3 R^2 + c_4 R^3 + c_5
R^4)\,\theta\left[V^-(R)-V_0(R)\right]\,, \label{fitting func}
\ee
where $c_1 = -302.66$~eV, $c_2 = 635.8$~eV/a$_0$, $c_3 = -480.06$~eV/a$_0^2$, $c_4 = 156.9$~eV/a$_0^3$, $c_5 = -18.88$~eV/a$_0^4$; $\theta$ is the step function. This fit and the calculated results are displayed in Fig.~\ref{gammaCO}. In the local version of complex-potential model is not possible to calculate the level shift $\Delta(R)$ from first principles (this requires $\Gamma(E,R)$). In the LCP approach the level shift is therefore treated a phenomenological parameter external to the model. We use this degree of freedom to match correctly the position of the peaks using the  $0\to1$ and $0\to2$ cross sections of Allan~\cite{PhysRevA.81.042706}. Moreover we  scale the width $\Gamma(R)$ to correctly reproduce the experimental height of the main peaks. In practice $\G(R)$ was multiplied by 1.1 and the $\D$ parameter was set to 0.035~eV. Fig.~\ref{fig:COxsec_allan} compares our  \emph{ab initio} cross sections and the final results.

Fig.~\ref{fig:COxsec_allan} compares our vibrational excitation cross sections for transition starting from the lowest vibrational level ($v_i=0$), with the measurements of Allan~\cite{PhysRevA.81.042706} and the previous {\it ab initio} predictions of Morgan~\cite{0953-4075-24-21-015}. It can be seen that the agreement between the three studies is very good for the lower vibrational levels. The discrepancy for high-lying levels, as shown by $0\to10$ even when  the experimental error of 20\%\ is taken into account, suggests a full non-local model and an energy-dependant widths, $\Gamma$, are needed for these states. However, we note that the cross sections for these extreme
excitations are very small.

\begin{figure}
%\begin{center}
\begin{indented}
\item[]
\includegraphics[scale=.7,angle=0]{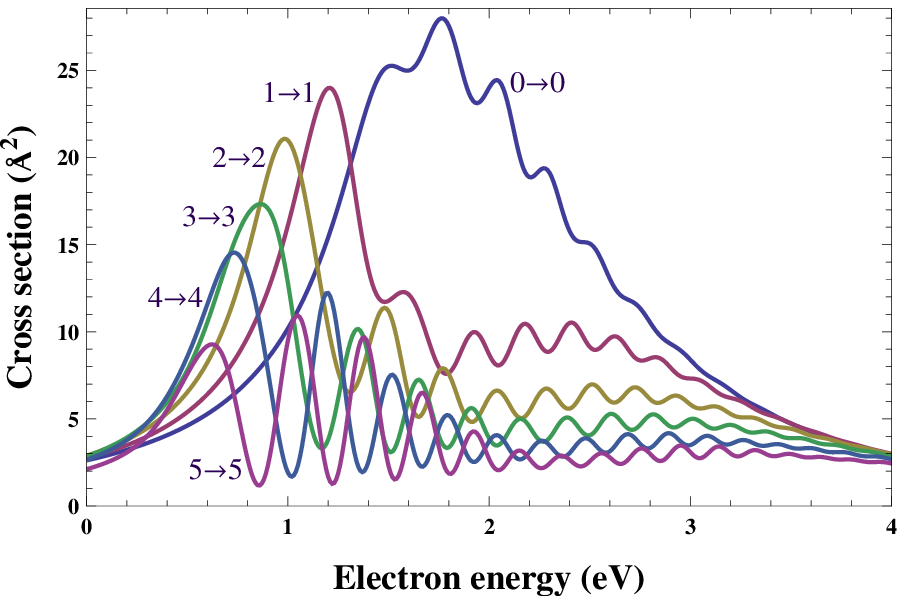}\hfill%\hspace{.5cm}
\includegraphics[scale=.7,angle=0]{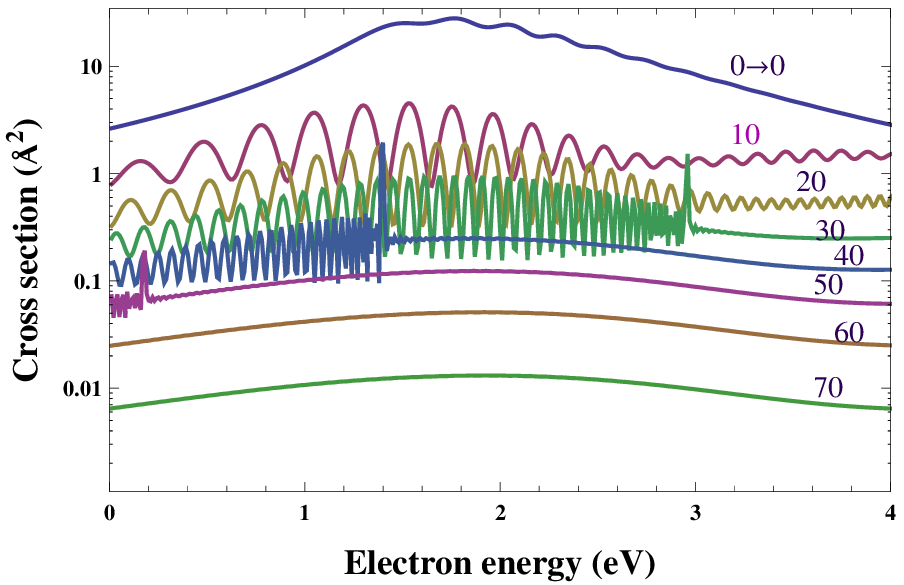}
\\
\vspace{.5cm}
\includegraphics[scale=.7,angle=0]{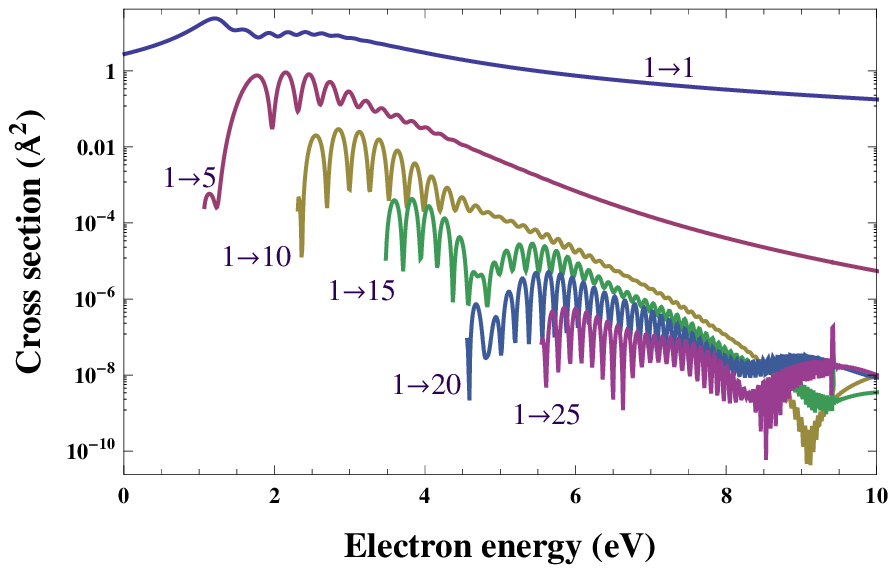}\hfill%\hspace{.5cm}
\includegraphics[scale=.7,angle=0]{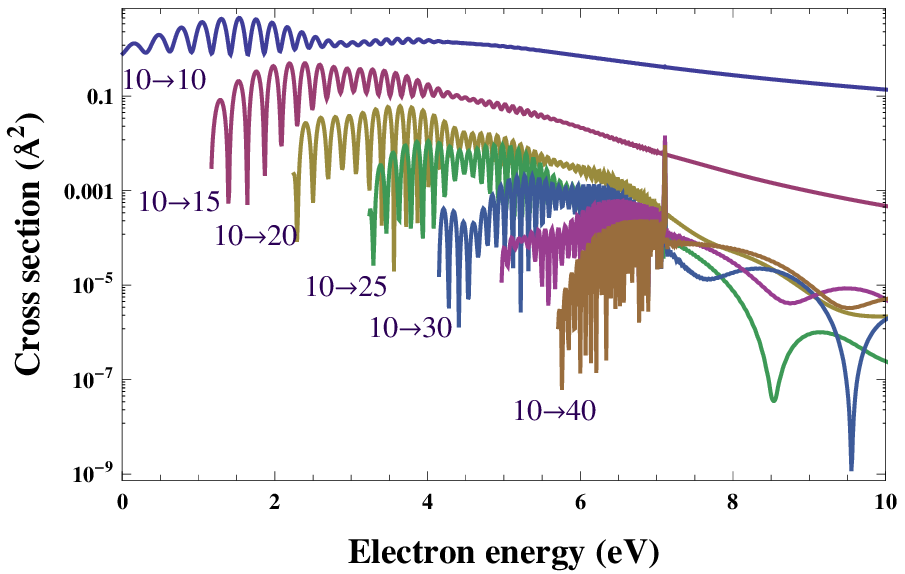}
%\end{center}
\end{indented}
\caption{e-CO resonant cross sections as a function of the incident electron energy. Upper panels: elastic processes involving the $v_i=v_f$ vibrational levels, as indicated in the figures. Lower panels: $v_i\rightarrow v_f$ processes starting from $v_i=1$ (left),  $v_i=10$ (right) and $v_f\geq v_i$. \label{fig:COxsec}}
\end{figure}

Having developed a satisfactory model for electron impact vibrational excitation of CO we  apply it to the whole range of vibrational states supported by the CO molecule. Figure~\ref{fig:COxsec} gives sample results showing that the excitation cross sections are strongly state dependent.

\begin{figure}
%\begin{center}
\begin{indented}
\item[]
\includegraphics[scale=.7,angle=0]{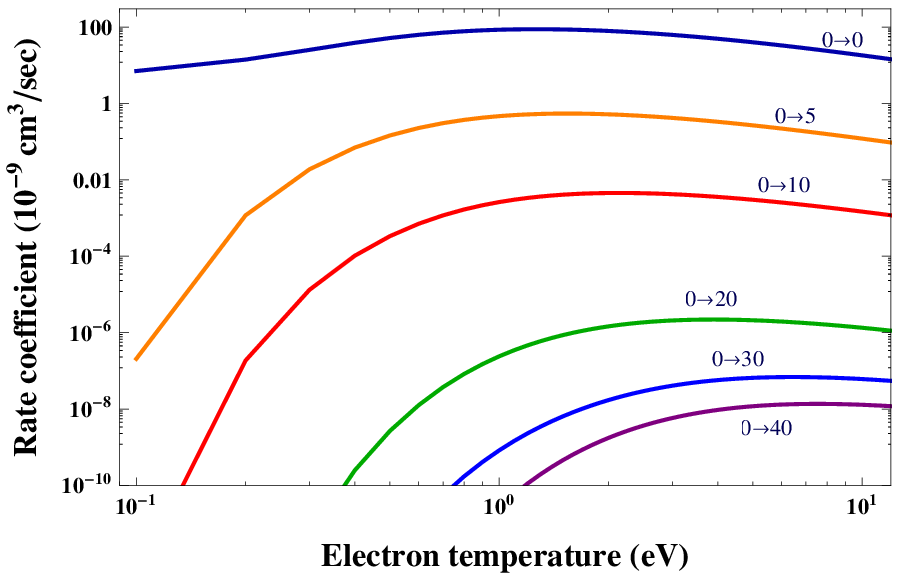}%\hspace{1cm}
\includegraphics[scale=.7,angle=0]{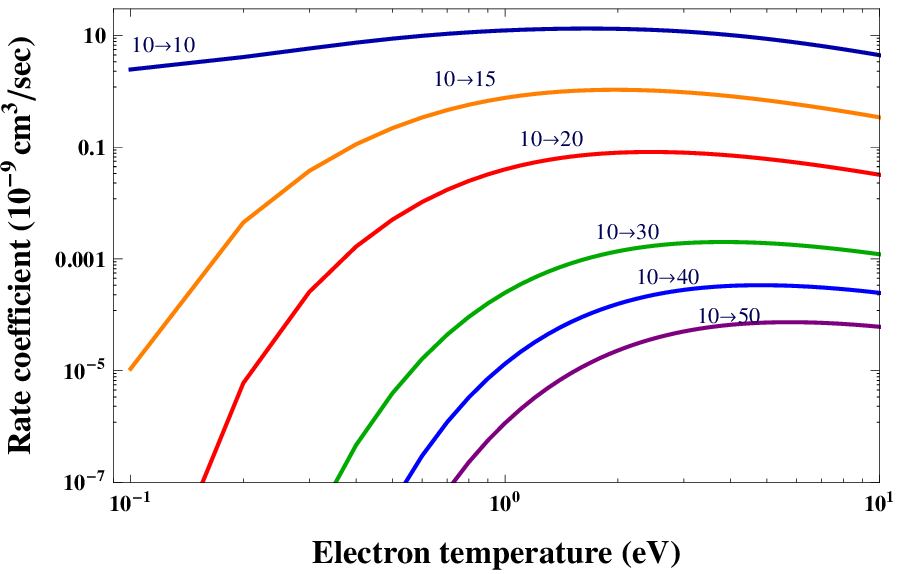}
%\end{center}
\end{indented}
\caption{e-CO resonant vibrational-inelastic excitation rate coefficients for the processes $0\rightarrow v_f$ (left panel) and $10\rightarrow v_f$ (right panel).\label{fig:vi_vf_rates}}
\end{figure}

Fig.~\ref{fig:vi_vf_rates} shows the rate coefficients for the inelastic transitions $v_i\rightarrow v_f$,  calculated from Eq.~(\ref{rate_coeff}) assuming a Maxwellian electron energy distribution function. The left panel of this figure illustrates the behavior of the rate coefficients for the inelastic processes starting from the level $v_i=0$. The decrease of the rates with $v_f$ can be attributed to the reduction in the corresponding cross sections, as is seen in Fig.~\ref{fig:COxsec_allan}. Analogous behavior is observed in the right panel, where the rates for vibrational excitations starting from $v_i=10$ are shown.

\begin{figure}
%\begin{center}
\begin{indented}
\item[]
\includegraphics[scale=.7,angle=0]{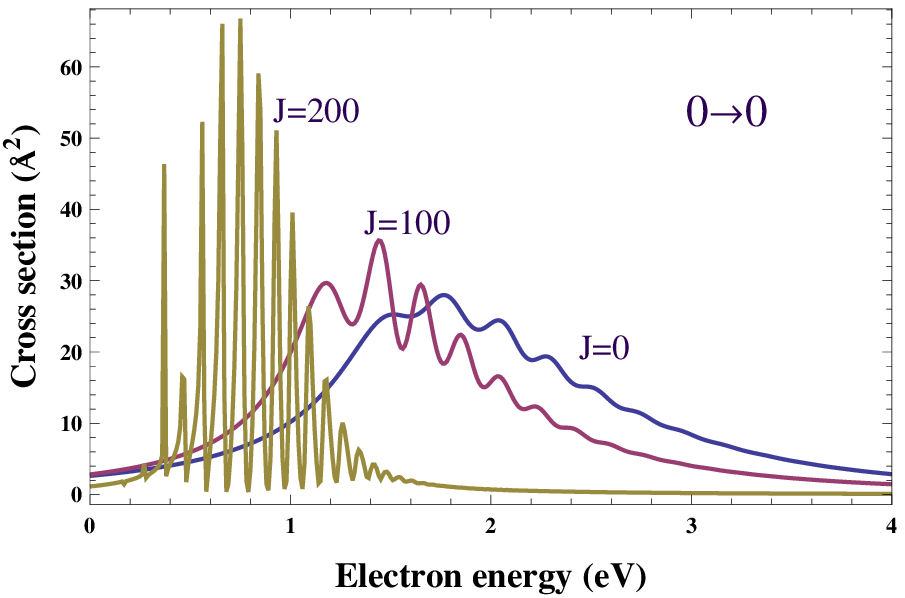}%\hspace{1cm}
\includegraphics[scale=.7,angle=0]{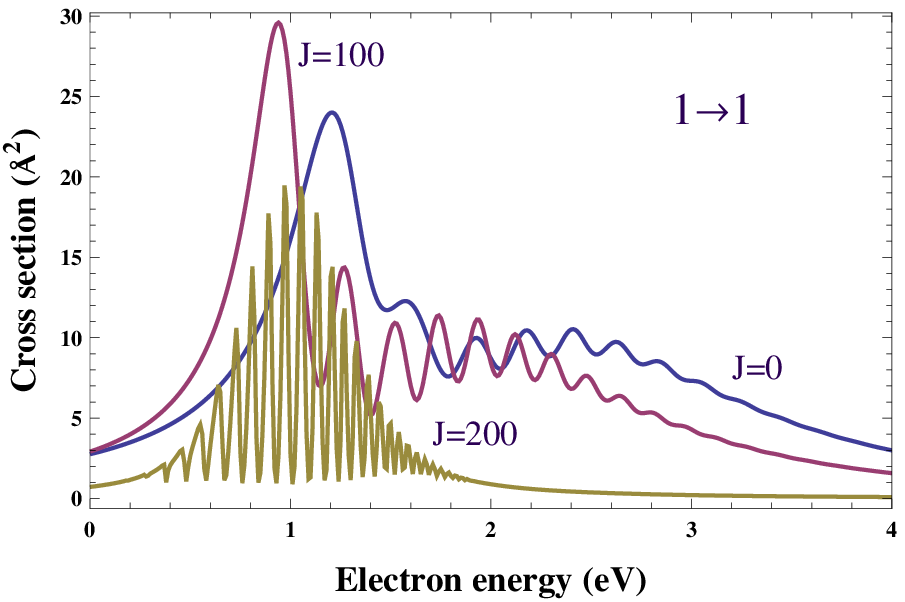}
\\%\vspace{.5cm}
\includegraphics[scale=.7,angle=0]{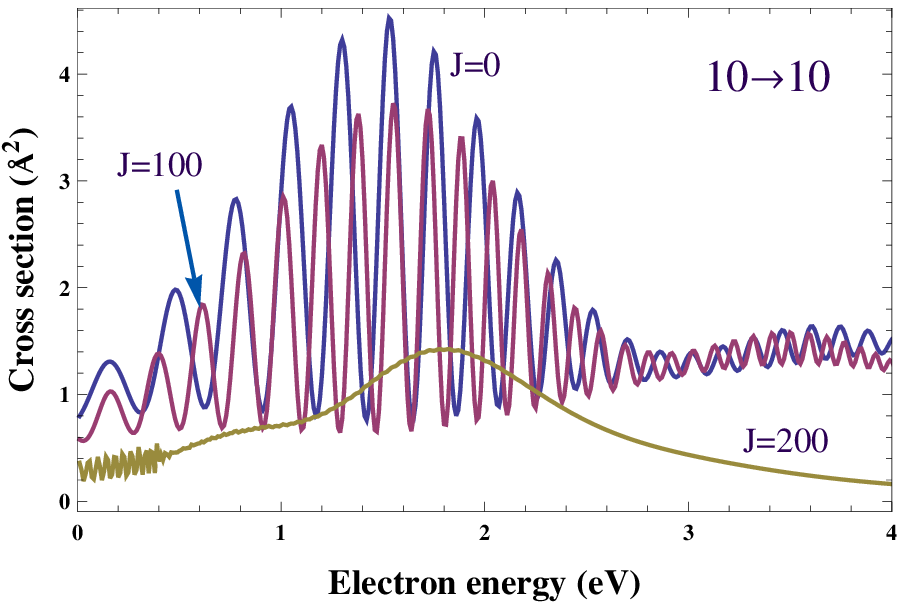}
%\end{center}
\end{indented}
\caption{e-CO resonant vibrational excitation cross sections calculated for different values of $J$. \label{fig:COxsec_J}}
\end{figure}

\begin{figure}
%\begin{center}
\begin{indented}
\item[]
\includegraphics[scale=.7,angle=0]{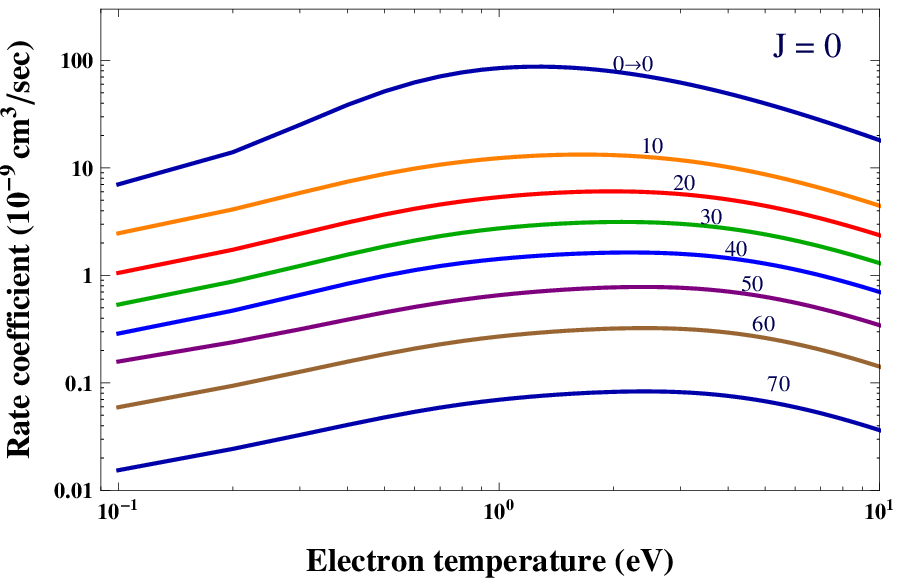}%\hspace{1cm}
\includegraphics[scale=.7,angle=0]{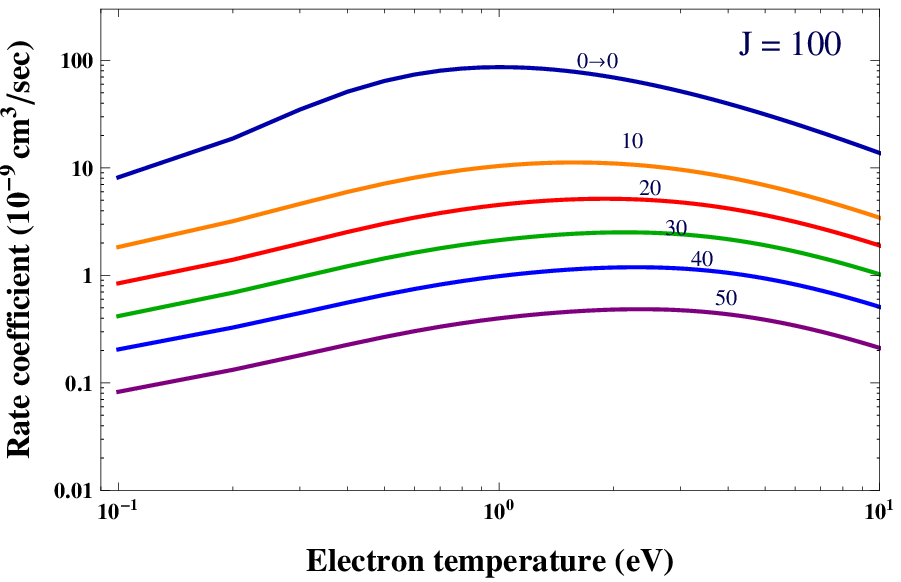}
\\%\vspace{.5cm}
\includegraphics[scale=.7,angle=0]{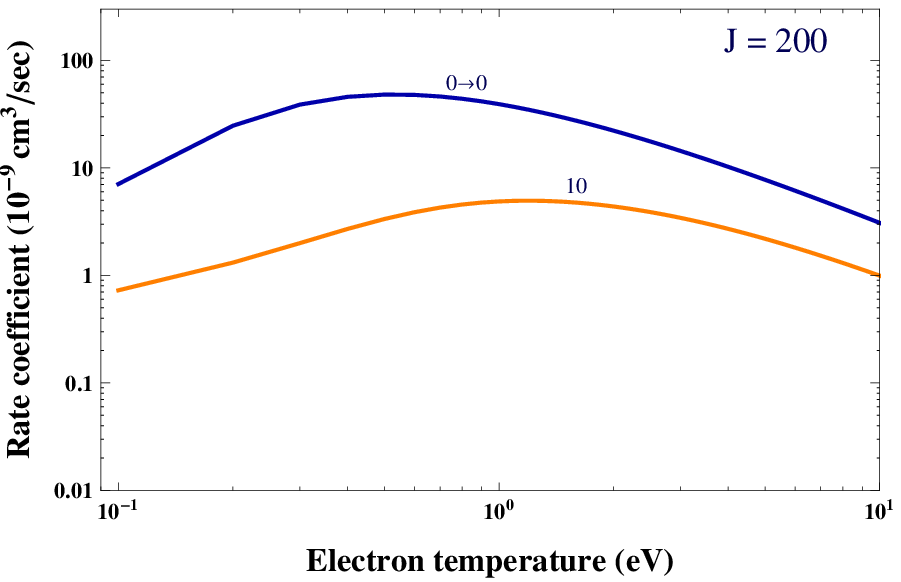}
%\end{center}
\end{indented}
\caption{Rate coefficients as a function of the electron temperature for e-CO elastic scattering calculated for $J=0,100$ and $200$. \label{fig:COrate_el_J}}
\end{figure}

We have also tested the effect of including rotational motion in the calculation. Figs.~\ref{fig:COxsec_J} and \ref{fig:COrate_el_J} show, respectively, some examples of cross sections and rates coefficients for different values of the angular momentum quantum number $J$. For cool CO samples, which only probe low-lying rotational states, the effect of including rotational motion is small. However for hot samples of CO, which occupy highly excited $J$ states, the magnitude, structure and position of the resonance enhanced excitation cross section changes significantly with rotational
excitation.

\section{Summary \label{sec:summary}}

The aim of this work is to provide electron impact excitation cross sections and rate coefficients of CO for modeling purposes. The theoretical cross section calculations have been performed in the framework of the local complex potential model for resonant collisions while the input parameters, adiabatic potential energies and widths, have been computed by the R-matrix method. The results obtained are in good agreement with the existing experimental and theoretical data.

A full set of cross sections and state-dependent rates can be obtained from~\cite{database}.

\ack
We wish to thank Dr. Annarita Laricchiuta and Dr. Domenico Bruno for helpful discussions.
This work was performed under the auspices of the European
Union Phys4Entry project funded under grant FP7-SPACE-2009-1 242311.

\section*{References}

\bibliographystyle{unsrt}{}

\end{document}